\documentstyle[aps,preprint,epsfig,feynmp]{revtex}

\def\be{\begin{eqnarray}}
\def\ee{\end{eqnarray}}

\newcommand{\bea}{\begin{eqnarray}}
\newcommand{\eea}{\end{eqnarray}}       

\tightenlines

\draft
\begin{document}

\title{Heavy quarkonium production in double Pomeron exchange processes
in perturbative QCD}

\author{Feng Yuan}
\address{Institut f\"ur Theoretische Physik, Universit\"at Heidelberg, 
Philosophenweg 19, 69120 Heidelberg, Germany}

\date{March 2001}
\maketitle

\begin{abstract}
We calculate the heavy quarkonium production in double pomeron
exchange processes in perturbative QCD by using two-gluon exchange model.
For the $P$-wave $\chi_J$ productions, we find $\chi_1$ and $\chi_2$ 
production amplitudes which vanish in the forward scattering limit.
We also calculate direct $J/\psi(\Upsilon)+\gamma$ production in 
the same approach, and these direct contributions
are much smaller than the feeddown contributions from the $P$-wave states.

\end{abstract}

In recent years, there has been a renaissance of interest in
diffractive scattering.
These diffractive processes are described by the Regge theory in
terms of the Pomeron ($I\!\! P$) exchange\cite{pomeron}.
The Pomeron carries quantum numbers of the vacuum, so it is a colorless entity
in QCD language, which may lead to the ``rapidity gap" events in experiments.
However, the nature of Pomeron and its reaction with hadrons remain a mystery.
For a long time it had been understood that the dynamics of the
``soft pomeron'' is deeply tied to confinement\cite{dosch}.
However, it has been realized now that much can be learned about
QCD from the wide variety of small-$x$ and hard diffractive processes,
which are now under study experimentally.
In Refs.\cite{th1,th2}, the diffractive $J/\psi$ and $\Upsilon$ production
cross section have been formulated in photoproduction processes and in
DIS processes in perturbative QCD.
In the framework of perturbative QCD the Pomeron is represented by a pair of
gluon in the color-singlet state.
This two-gluon exchange model can successfully describe the experimental
results from HERA\cite{hera-ex}.

This two-gluon exchange model has been extended to the hadron hadron 
collisions for the single 
diffractive processes\cite{levin,sd-yuan},
where the so-called coherent diffraction mechanism \cite{cfs}
plays an important role.
In these processes, the Pomeron represented by a color-singlet two-gluon
system is emitted from one hadron and interacts with another hadron in a hard
process, in which the two gluons are both involved.
These coherent diffraction processes are very interesting for the
study of hard diffraction mechanism in hadron collisions, where
there exist nonfactorization effects in the leading twist\cite{cfs}.

In this paper, we will further extend the above ideas to the double 
Pomeron exchange (DPE) processes. 
In particular, we will calculate heavy quarkonium 
productions in DPE processes at hadron colliders.
Apart from the single diffractive processes, the 
DPE processes at $p\bar p$ or $pp$ colliders are also
useful to the studies of the diffraction mechanism and exploration of the 
nature of the Pomeron.
In the DPE, both incoming hadrons are quasielastically scattered and
therefore emerge at very large Feynman $x_F$,
$
\label{xf} 
\mbox{$\vert$}
x_{F,1}
\mbox{$\vert$}
,~
\mbox{$\vert$}
x_{F,2}
\mbox{$\vert$}
\ge 0.9.
$
And the final states system in the center region
is produced from the Pomeron-Pomeron interaction.
Due to the
color-singlet (Pomeron) exchange, the final states system can be separated
from the two beam particles (proton or antiproton) by large rapidity gaps.
These DPE processes are quite attractive because it has been long time
proposed to search Higgs meson in double rapidity gap events at proton 
colliders\cite{h1,h2,h3,h4,h5,h6,h7}. 
An obvious advantage of the rapidity gap approach is the 
spectacularly clean experimental signature\cite{h8}. 

In the two-gluon exchange model, 
$\chi_c$ and $\chi_b$ states productions in DPE processes have been estimated 
in Ref.\cite{martin}, where they approximately take the
gluons coupling to the $\chi$ states in $gg\rightarrow \chi_J$ from  
their decay widths in $\chi_J\rightarrow gg$ processes.
In this paper, we will directly calculate these amplitudes in terms of 
the wave functions derivatives at the origin of the $P$-wave heavy 
quarkonium bound states in the framework of color-singlet model by using the
nonrelativistic approximation, 
and we also take into account the strong correlations between the two
incident gluons polarizations, which will enter into the production amplitudes
explictly.
As a result, we find vanishing amplitudes for $\chi_1$ and $\chi_2$ states
in the forward scattering limit.
We will also calculate $J/\psi+\gamma$ production in DPE processes
in the same approach.
  
The double pomeron exchange processes $\chi$ production in
\begin{equation}
\label{process}
p+p(\bar p)\rightarrow p+\chi_J+p(\bar p)
\end{equation}
is plotted in Fig.~1
in the two gluon exchange model, where the hard 
subprocess $gg\rightarrow M$ is initiated by gluon-gluon fusion and the 
second $t$-channel gluon (with transverse momentum $Q_T$) 
is needed to screen the color flow across the rapidity gap intervals. 
In our calculations, we express the formulas in terms of the Sudakov
variables.
That is, every four-momenta $k_i$ are decomposed as,
\begin{equation}
k_i=\alpha_i p_1+\beta_i p_2+\vec{k}_{iT},
\end{equation}
where $p_1$ and $p_2$ are the momenta of the incident hadrons (proton or
antiproton). For high energy process, we will have 
$p_1^2=0$, $p_2^2=0$, and $2p_1\cdot p_2=s$.
$\alpha_i$ and $\beta_i$ are the momentum fractions of $p_1$ and $p_2$
respectively.
$k_{iT}$ is the transverse momentum, which satisfies
\begin{equation}
k_{iT}\cdot p_1=0,~~~
k_{iT}\cdot p_2=0.
\end{equation}
For the momenta of the incident gluons $q_1$ and $q_2$, we
have the following decompositions,
\begin{equation}
q_1=x_1 p_1+q_{1T},~~~~q_2=x_2 p_2+q_{2T}.
\end{equation}
And then, the momentum of the produced $\chi$ state will have the following 
form,
\begin{equation}
P(\chi)=q_1+q_2=x_1 p_1+x_2 p_2+P_{T},
\end{equation}
where the $\chi$ state transverse momentum 
$\vec{P}_T=\vec{q}_{1T}+\vec{q}_{2T}$.
In the following, we will calculate the amplitudes in the forward scattering 
limit. That is to say, the momenta transfer squared between the final state
hadrons and initiate hadrons are set to be zero,
$t_1=(p_1-p_1')^2=0$ and $t_2=(p_2-p_2')^2=0$. 
Under this limit, we will have 
$\vec{q}_{1T}=-\vec{q}_{2T}=\vec{Q}_{T}$, and then $\vec{P}_T(\chi)=0$.
With the above kinematical considerations, 
we can formulate the forward scattering amplitudes
for $pp\rightarrow p+\chi+p$ as follows,
\begin{equation}
\label{amp}
{\cal M}(\chi_J)=\frac{9\pi^3}{4}\int \frac{d^2Q_T}{\pi(Q_T^2)^2}
	f_g(x_1,x_1',Q_T,M)f_g(x_2,x_2',Q_T,M)\times A_J,
\end{equation}
where $f_g$ are the unintegrated off-forward (skewed) gluon distribution
functions. They can be related to the conventional gluon density as
\cite{martin}
\begin{equation}
f_g(x,x',Q_T,M)=R_g\frac{\partial}{\partial {\rm ln} Q_T^2}
[\sqrt{T(Q_T,M)}xg(x,Q_T^2)].
\end{equation}
The factor $R_g$ takes into account the skewed effects of the off-forward
gluon density compared with the conventional gluon density in the region of
$x'\ll x$. The bremsstrahlung survival 
probability $T^2$ can be found in \cite{martin}, which will reduce to 
the conventional Sudakov form factors in the double logarithmic limit.

The expressions $A_J$ in the amplitude (\ref{amp}) include the $\chi_J$
bound state effects (the wave functions), and will also
depend on the polarizations of $\chi_J$ states. 
For $J=0$, we have the following simple result,
\begin{equation}
A_0=\frac{128\pi\alpha_s M}{9(M^2+Q_T^2)^2}
	\sqrt{\frac{1}{4\pi M}}R'_P(0),
\end{equation}
where $R'_P(0)$ is the radial wave function derivative at the origin for
the $P$-wave $\chi$ state of charmonium or bottomonium.
For $J=1$, 
\begin{equation}
\label{a1}
A_1=\frac{8\times 128\pi\alpha_s }{9\sqrt{6}s(M^2+Q_T^2)^2}
 \sqrt{\frac{1}{4\pi M}}R'_P(0) \epsilon_{\mu\nu\rho\sigma}
	e_{(J=1)}^\mu Q_T^\nu p_1^\rho p_2^\sigma,
\end{equation}
where $e_{(J=1)}$ is the polarization vector of $\chi_1$ state.
From the above expression we can see that $A_1\propto Q_T^\nu$,
so the amplitude for $J=1$ state will be zero after angular integration of
(\ref{amp}) over $\vec{Q}_T$.
That is to say, there is 
no contribution for $\chi_1$ production in the DPE processes
(\ref{process}).

For $J=2$, 
\begin{equation}
\label{a2}
A_2=\frac{64\pi\alpha_s M}{9\sqrt{3}sQ_T^2(M^2+Q_T^2)^2}
	\sqrt{\frac{1}{4\pi M}}R'_P(0)e^{\mu\nu}_{(J=2)}
[4Q_T^2(p_{1\mu} p_{2\nu}+p_{1\nu} p_{2\mu})+s(P_\mu P_\nu-
4Q_{T\mu} Q_{T\nu})].
\end{equation}
Here $e^{\mu\nu}_{(J=2)}$ is the polarization tensor for $\chi_2$ state,
which has the following properties,
\begin{equation}
\label{e2}
e_{\mu\nu}P^\mu=0,~~e_{\mu\nu}g^{\mu\nu}=0.
\end{equation}
For the angular integration of (\ref{amp}) over $\vec{Q}_T$, 
we have the following identity,
\begin{equation}
\int d^2Q_T Q_T^\mu Q_T^\nu=\frac{\pi}{2}\int dQ_T^2 Q_T^2 g_{\mu\nu}^{(T)},
\end{equation}
where the transverse part of $g_{\mu\nu}$ is given by
$$g_{\mu\nu}^{(T)}=-g_{\mu\nu}+\frac{2}{s}(p_{1\mu}p_{2\nu}+p_{1\nu}p_{2\mu})
$$
So, after integrating the azimuth angle of $\vec{Q}_T$, 
the expression in the brackets $[~~]$ of Eq.~(\ref{a2})
will become
\begin{equation}
[~~]=s(P_\mu P_\nu-2g_{\mu\nu}Q_T^2).
\end{equation}
With the above results, we find that the amplitude for $\chi_2$ will be
equal to zero because of the identities of Eq.~(\ref{e2}).

The vanishing of $\chi_1$ and $\chi_2$ production amplitudes 
are direct consequences of the forward scattering limits in the
DPE processes formulated in Eq.~(\ref{amp}).
Under this limit, the polarization vectors of the two incident gluons are
strongly correlated, i.e., $\vec{e}_i\propto \vec{q}_{iT}$ with 
$\vec{q}_{1T}=-\vec{q}_{2T}$\cite{pumplin,berera,khoze,martin}.
Because of these strong correlations, the final state ($\chi_J$ state here)
in the hard DPE processes must be in the $J_z=0$ state\cite{khoze}.
On the other hand, we know that the ${}^3P_2$ state of $J_z=0$
decouples with two real gluons in
the nonrelativistic limit for the heavy quarkonium system
\cite{alekeev}. 
By explicit calculations in the above, we show that 
the DPE amplitudes for $\chi_1$ and $\chi_2$
productions have the properties in (\ref{a1}) and (\ref{a2}), 
and finally give vanishing contribution to these states.
We note that these properties are common results in the two-gluon exchange
Pomeron model (either perturbative or nonperturbative two-gluon exchange
models).
However, in the other Pomeron model, such as D-L model\cite{dl}, 
the production amplitudes for $\chi_1$ and $\chi_2$ do not vanish
\cite{schafer,peng}. 
This is because in this model the Pomeron couples approximately 
like a $C=+1$ photon, and the DPE $\chi_J$ amplitudes were adopted 
from the analogous process of $\chi_J\rightarrow \gamma^*\gamma^*$,
which will give the same magnitude production rates for $\chi_1$ and 
$\chi_2$ as that of $\chi_0$\cite{schafer,peng}.

Finally, the differential cross section of $\chi_0$ production
will be
\begin{equation}
d\sigma=\frac{|{\cal M}|^2}{16^2\pi^3}e^{bt_1}e^{bt_2}dt_1dt_2dy,
\end{equation}
where $y$ is the rapidity of $\chi_0$ state. 
$t_i$ is the momentum transfer squared at the proton (antiproton) vertices, 
and $b$ is slope for the proton form factor, which will be taken as 
$b=4.0GeV^{-2}$ in the following numerical calculations.
After integrating $t_1$ and $t_2$, we get
\begin{eqnarray}
\label{xs-ch}
\frac{d\sigma}{dy}&=&\frac{|{\cal M}|^2}{16^2\pi^3 b^2}=
\frac{12\pi^4\alpha_s^2 M}{b^2}|R'_P(0)|^2\left [\int \frac{dQ_T^2}{(Q_T^2)^2}
\frac{f_g(x_1,x_1',Q_T,M)f_g(x_2,x_2',Q_T,M)}{(M^2+2Q_T^2)^2}\right ]^2.
\end{eqnarray}

In Fig.~2, we plot the differential cross sections (\ref{xs-ch}) for
$\chi_{c0}$ and $\chi_{b0}(1P)$ at the Fermilab Tevatron RUN II, 
where we adopt the rapidity gap survival probability to be $S^2=0.05$ 
as in Ref.~\cite{martin}.
For the masses and wave functions\cite{potential}, we use
\begin{eqnarray}
\nonumber
M(\chi_{c0})=3.4 GeV, ~~|R'_P(0)|^2_c=0.075GeV^5,\\
M(\chi_{b0})=10.2 GeV, ~~|R'_P(0)|^2_b=1.42GeV^5.
\nonumber
\end{eqnarray}
For the gluon distribution function, we use the GRV LO parameterization
\cite{grv}.
After integrating over the rapidity, we get the following cross sections for
double-diffractive productions of $chi_{c0}$ and $\chi_{b0}$ as
\begin{equation}
\sigma(\chi_{c0})=735~{\rm nb},~~ 
\sigma(\chi_{b0})=0.88~{\rm nb}
\end{equation}.
Comparing our results with those of \cite{martin}, we find that 
our result for $\chi_{c0}$ is roughly agreeing with \cite{martin}. However,
our result  for $\chi_{b0}$ 
is much larger than that in\cite{martin}. This is because we use the
potential model calculations for the wave functions, while they use lattice 
calculations for the decay widths for $\chi_{b}$ states.
So, for $b\bar b$ bound states, there exist large theoretical uncertainties
associated with the bound states properties.

For the experimental observation, with the high resolution missing-mass 
measurements at the Tevatron ($\Delta M\approx 250MeV$\cite{h8}), 
we can identify all of the $\chi$ states in
the DPE processes. 
Furthermore, we can also measure $\chi_0$ productions in DPE by 
detecting their radiative
decays to $J/\psi$ ($\Upsilon$) plus photon. 
The branching ratio of 
$\chi_0\rightarrow J/\psi \gamma$ is about $6.6\times 10^{-3}$,
so the final cross section for $J/\psi$ production from $\chi_{c0}$ feeddown 
decays is about 
$\sigma(\chi_{c0})\times Br(\chi_{c0}\rightarrow J/\psi\gamma)\times 
Br(J/\psi\rightarrow \mu^+\mu^-)\approx 300~{\rm pb}$.
Having this large cross section, 
$J/\psi$ production in the DPE processes can be
well measured with large number of lepton pair ($\mu^+\nu^-$ or $e^+e^-$)
events.
For $P$-wave bottomonium states, we know that the branching ratio of 
$\chi_{b0}(1P)\rightarrow \Upsilon(1S)\gamma)$ is very small (the upper bound
is $BR<6\%$), so it may be difficult to detect $\Upsilon(1S)$ from 
$\chi_{b0}(1P)$ decays. 
However, for $2P$ state $\chi_{b0}(2P)$, we have not so small branching
ratios decaying into $\Upsilon(1S)$ and $\Upsilon(2S)$.
So, from the combining branching ratios, we will have
\begin{eqnarray}
\nonumber
\sigma(\chi_{b0}(2P))&\times & \left (Br(\chi_{b0}(2P)\rightarrow 
\Upsilon(2S)\gamma\rightarrow \mu^+\mu^-)+
Br(\chi_{b0}(2P)\rightarrow 
\Upsilon(1S)X\rightarrow \mu^+\mu^-)\right )\\
&\approx & 0.8~{\rm pb}.
\nonumber
\end{eqnarray}
In the above estimate, we assumed that the production cross section of 
$\chi_{b0}(2P)$ is the same as that of $\chi_{b0}(1P)$ because their wave
functions at the origin are roughly the same\cite{potential}.
So, $\Upsilon$ states from the feeddown decays of $\chi_{b0}(2P)$
can also be observable in the DPE processes 
at the Tevatron II with integrated luminosity of $15{\rm fb}^{-1}$.

Apart from the feeddown contributions from $\chi_J$ states, 
$J/\psi$ and $\Upsilon$ can also be produced directly from the 
DPE processes in 
\begin{equation}
p+p(\bar p) \rightarrow p+J/\psi\gamma+p(\bar p),
\end{equation}
for which we plot in Fig.~3 the typical diagram. 
For the final produced $J/\psi$, we will have the Sudakov variables as
$$P=\alpha_\psi p_1+\beta_\psi p_2+P_T,$$
and for the outgoing photon,
$$p_\gamma=\alpha_3 p_1+\beta_3 p_2+p^\gamma_T.$$
In the forward scattering limit, because $\vec{q}_{1T}+\vec{q}_{2T}=0$,
the transverse momentum of the photon will be equal to the transverse
momentum of $J/\psi$ in balance, i.e., $\vec{p}^\gamma_T+\vec{P}_T=0$. 
The onshell conditions for the outgoing  
$J/\psi$ and photon will give the following relations for the associated
Sudakov variables,
\begin{equation}
\alpha_\psi\beta_\psi=\frac{P_T^2+m^2}{s}\equiv \frac{m_T^2}{s},
~~\alpha_3\beta_3=\frac{P_T^2}{s}.
\end{equation}
From the momenta conservation, we also have
$$x_1=\alpha_\psi+\alpha_3,~~x_2=\beta_\psi+\beta_3.$$

Using the above Sudakov variables, we derive the differential cross section for
$J/\psi\gamma$ production in the DPE processes,
\begin{equation}
\frac{d\sigma}{dy_\gamma dy_\psi d^2 P_T}=\frac{|{\cal M}|^2}{16^3\pi^6 b^2},
\end{equation}
where $y_\psi$ and $y_\gamma$ are the rapidities of final state
$J/\psi$ and photon, and $\alpha_\psi$ and $\alpha_3$ can be related to
rapidities,
$$\alpha_\psi=\frac{m_T}{\sqrt{s}}e^{y_\psi},~~\alpha_3=
\frac{P_T}{\sqrt{s}}e^{y_\gamma}.
$$
The amplitude for $J/\psi+\gamma$ production is more complicated than 
those for $\chi_J$ productions.
In the limit of $Q_T^2\rightarrow 0$, we can simplify the amplitude and
get the amplitude squared as
\begin{equation}
|{\cal M}|^2=\frac{2\times 16^3\pi^8\alpha_s^2\alpha_e e_c^2\alpha_3^2
\alpha_\psi^2 m |R_S(0)|^2}{3x_1^2
(\alpha_3m_T^2+\alpha_\psi P_T^2)^2(m_T^2)^2}{\cal I}_g^2,
\end{equation}
where 
\begin{equation}
{\cal I}_g=\int \frac{dQ_T^2}{(Q_T^2)^2}
f_g(x_1,x_1',Q_T,M)f_g(x_2,x_2',Q_T,M).
\end{equation}
So the cross section will be
\begin{equation}
\frac{d\sigma}{dy_\gamma dy_\psi d^2 P_T}=
\frac{2\pi^2\alpha_s^2\alpha_e e_c^2\alpha_3^2
\alpha_\psi^2 m |R_S(0)|^2}{3b^2x_1^2
(\alpha_3m_T^2+\alpha_\psi P_T^2)^2(m_T^2)^2}{\cal I}_g^2.
\end{equation}

In Fig.~4, we plot the differential cross sections $d\sigma/dydP_T$ as
functions of $P_T$ at $y=0$ for $J/\psi$ and $\Upsilon(1S)$ at the Tevatron
RUN II. For the numerical inputs for masses and wave functions, 
we use\cite{potential}
\begin{eqnarray}
\nonumber
M(J/\psi)=3.1 GeV, ~~|R_S(0)|^2_c=0.81GeV^5,\\
M(\Upsilon(1S))=9.46 GeV, ~~|R_S(0)|^2_b=6.48GeV^5.
\nonumber
\end{eqnarray}
From Fig.~4, we see that the cross section for $J/\psi$ drops very rapidly 
with $P_T$, and the cross section for $\Upsilon$ drops much more 
slowly than that of $J/\psi$. 
At low $P_T$, the cross section of $J/\psi$ is much larger
than that of $\Upsilon$, but at large enough $P_T$
it will be smaller than $\Upsilon$.
However,
comparing the results in Fig.~4 to those in Fig.~2, we find that the cross 
sections for $S$-wave quarkonia are much smaller than those for $P$-wave 
states, even after the latter cross sections have been multiplied by the decay 
branching ratios of $\chi\rightarrow J/\psi (\Upsilon) X$. 
So, we conclude that the 
dominant contributions to $S$-wave quarkonium production in the DPE processes
are the feeddown contributions from $P$-wave decays. This is quite different
from the inclusive production processes or the DPE $J/\psi$ production in 
the IS Pomeron model\cite{yf-dpe}, where the direct production is the dominant
source to the prompt $J/\psi$ productions.

In conclusion, in this paper we have investigated the heavy quarkonium 
productions in the DPE processes at hadron colliders in perturbative 
QCD. 
We found that in the 
forward scattering limit, the amplitudes for $\chi_1$ and $\chi_2$ vanish
due to the strong correlation between the polarization vectors of the two
incident gluons.
We also calculated the direct $S$-wave quarkonium production in the DPE 
processes associated with a photon, and found that the direct contributions
are much smaller than the feeddown contributions from the $P$-wave states.

\acknowledgments
We thank V.A. Khoze and M.G. Ryskin for correspondence and useful 
conversations and discussions.
We are also grateful for the discussions with K.T. Chao, H.J. Pirner,
 and J.S. Xu.

\newpage
\centerline{\bf \large Figure Captions} \vskip 1cm \noindent
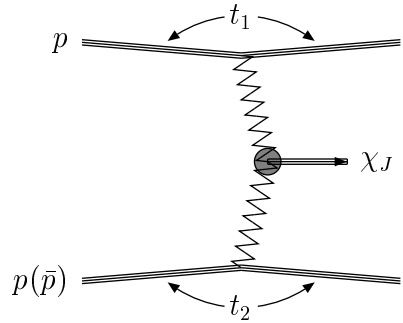
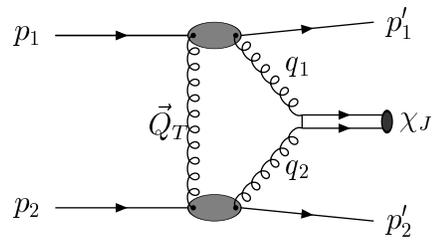
Fig.1. Sketch diagram of the DPE processes $\chi_J$ production at hadron 
colliders: (a) Pomeron-Pomeron fusion into $\chi_J$; (b) Double Pomeron
exchange in the perturbative two-gluon exchange picture.

\noindent
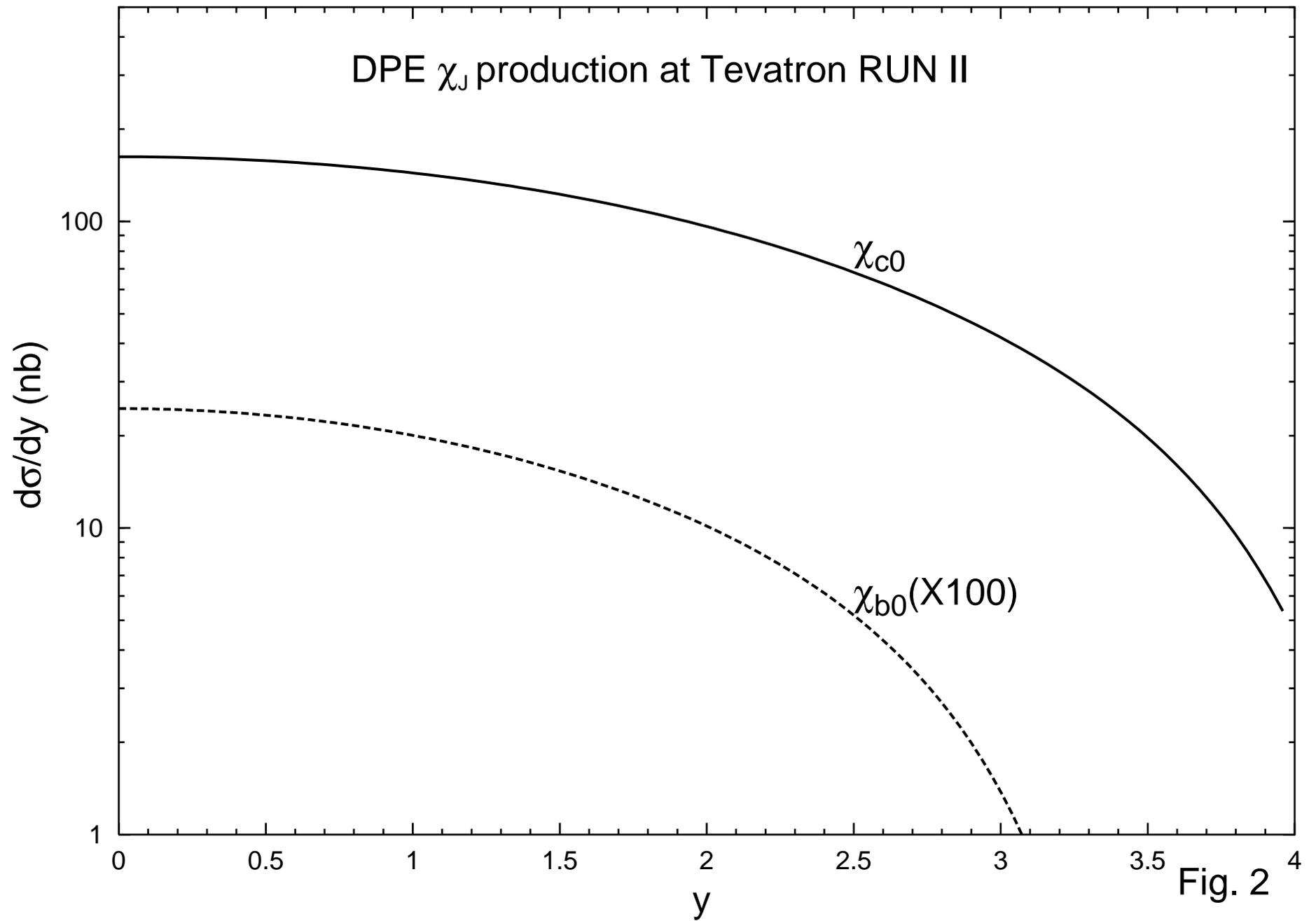
Fig.2. The differential cross sections $d\sigma/y$ for $\chi_{c0}$ and
$\chi_{b0}$ in DPE processes at the Fermilab Tevatron RUN II.

\noindent
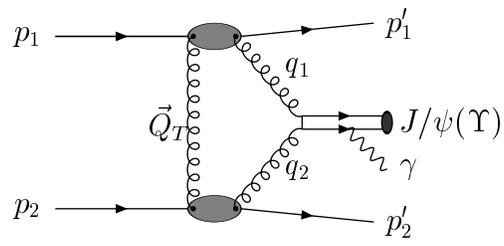
Fig.3. A typical diagram for the direct $J/\psi$ ($\Upsilon$) 
production in the DPE processes.

\noindent
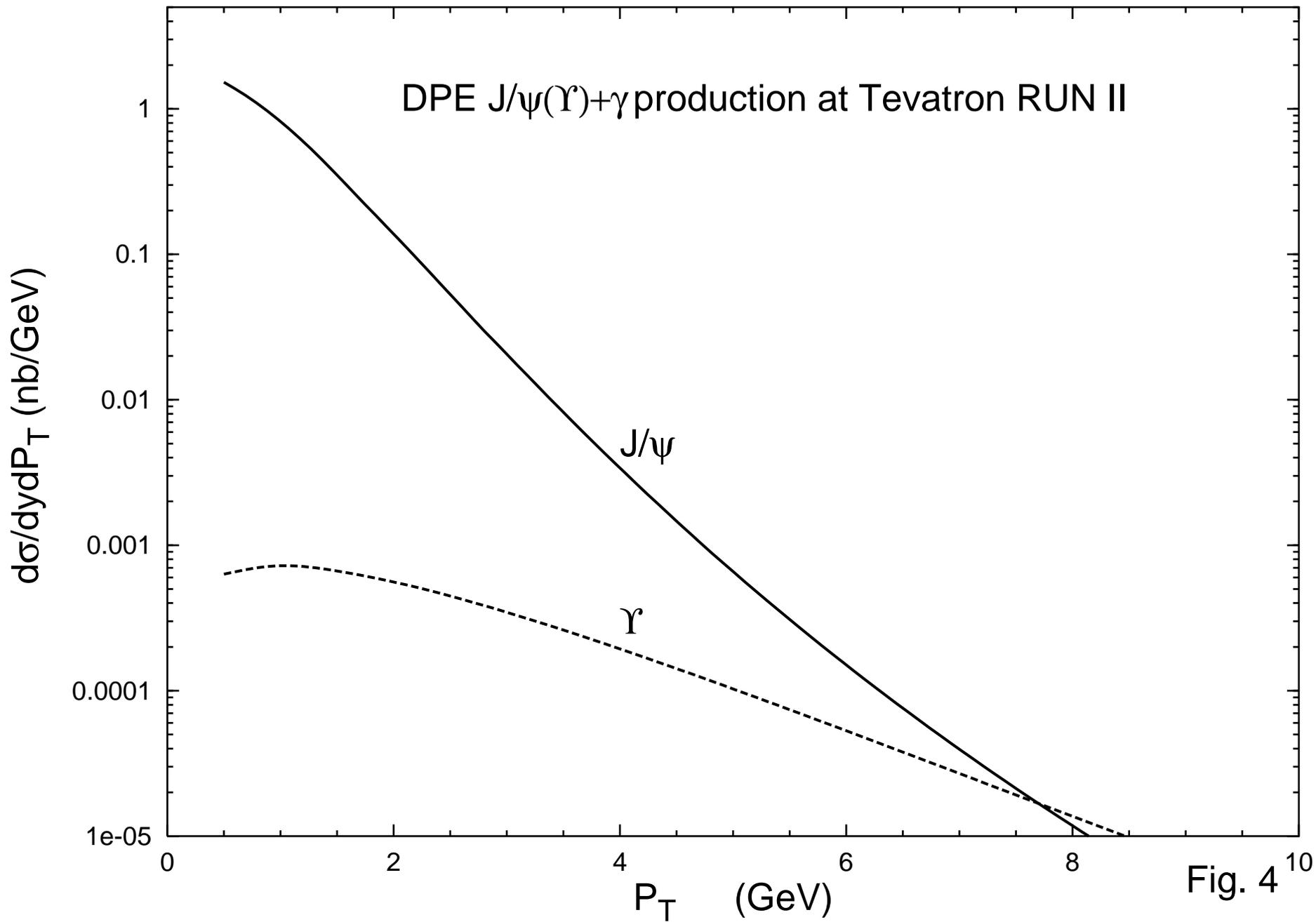
Fig.4. The differential cross sections $d\sigma/dydp_T$ for $J/\psi$ and
$\Upsilon$ production in the DPE at the Tevatron RUN II.

\end{document}